\documentclass[english,aps,nofootinbib]{revtex4}%
\usepackage[T1]{fontenc}
\usepackage[latin9]{inputenc}
\usepackage{amsmath}
\usepackage{color}
\usepackage{graphicx}
\usepackage{amssymb}
\usepackage{color}
\usepackage{color}
\usepackage{color}
\usepackage{color}
\usepackage{amsfonts}
\usepackage{babel}%
\makeatletter
\providecommand{\tabularnewline}{\\}
\@ifundefined{definecolor}
{
}{}
\makeatletter
\@ifundefined{definecolor}
{\@ifundefined{definecolor}
{
}{}
}{}
\makeatletter
\@ifundefined{definecolor}
{\@ifundefined{definecolor}
{\@ifundefined{definecolor}
{
}{}
}{}
}{}
\makeatletter
\@ifundefined{definecolor}
{\@ifundefined{definecolor}
{\@ifundefined{definecolor}
{
}{}
}{}
}{}
{}
\definecolor{blue}{rgb}{0.2,0.3,0.8}
\definecolor{red}{rgb}{0.8,0.2,0.2}
\makeatother
\makeatother
\makeatother
\makeatother
\begin{document}
\title{Nonlocal chiral quark models with wavefunction renormalization: sigma
properties and $\pi-\pi$ scattering parameters}
\author{S. Noguera$^{a}$}
\email{Santiago.Noguera@uv.es}
\author{N.N. Scoccola$^{b,c,d}$}
\email{scoccola@tandar.cnea.gov.ar}
\affiliation{$^{a}$ Departamento de Fisica Teorica and Instituto de F\i{}\'{s}ica
Corpuscular, Universidad de Valencia-CSIC, E-46100 Burjassot (Valencia), Spain.}
\affiliation{$^{b}$ CONICET, Rivadavia 1917, 1033 Buenos Aires, Argentina}
\affiliation{$^{c}$ Physics Department, Comisi\'{o}n Nacional de Energ\'{\i}a At\'{o}mica,
Av.\ Libertador 8250, 1429 Buenos Aires, Argentina }
\affiliation{$^{d}$ Universidad Favaloro, Sol{\'{\i}}s 453, 1078 Buenos Aires, Argentina}

\begin{abstract}
We analyze the sigma meson mass and width together with the pion-pion
scattering parameters in the context of non-local chiral quark models with
wave-function renormalization (WFR). We consider both non-local interactions
based on the frequently used exponential form factor, and on fits to the quark
mass and renormalization functions obtained in lattice calculations. In the
case of the sigma properties we obtain results which are less dependent on the
parameterization than in the standard local NJL model, and which are in
reasonable agreement with the recently reported empirical values. We also show
that the inclusion of the WFR tend to improve the description of the $\pi
$-$\pi$ scattering parameters, with the lattice inspired parameterization
providing the best overall results. Finally, we analyze the connection of the
non-local quark models discussed here with Chiral Perturbation Theory, and
present the model predictions for the low energy constants relevant for $\pi
$-$\pi$ scattering to $O(4)$ in the chiral expansion.

\end{abstract}

\maketitle


\renewcommand{\thefootnote}{\arabic{footnote}} \setcounter{footnote}{0}

\section{Introduction}

Although much effort has been made in trying to predict low energy hadron
observables directly from QCD, one is still far from reaching this goal due to
the extremely complex non-perturbative behavior of the theory in that regime.
In such a situation it proves convenient to turn to the study of effective
models. For two light flavors it is believed that QCD supports an approximate
SU(2) chiral symmetry which is dynamically broken at low energies, and pions
play the role of the corresponding Goldstone bosons. A simple scheme including
these properties is the well known Nambu$-$Jona-Lasinio (NJL)
model~\citep{Nambu:1961tp}, proposed more than four decades ago. The NJL model
has been widely used as an schematic effective theory for
QCD~\citep{Vogl:1991qt,Klevansky:1992qe,Hatsuda:1994pi}, allowing e.g.\ the
description of light mesons as fermion-antifermion composite states. In the
NJL model quarks interact through a local, chiral invariant four-fermion
coupling. Because of the local nature of this interaction, the corresponding
Schwinger-Dyson and Bethe-Salpeter equations become relatively simplified.
However, the main drawbacks of the model are direct consequences of this
locality: loop integrals are divergent (and therefore have to be regulated
somehow), and the model is nonconfining. As a way to improve upon the NJL
model, extensions which include nonlocal interactions have been proposed (see
Ref.~\citep{Rip97} and references therein). In fact, nonlocality arises
naturally in quantum field theory and, particularly, in several well
established approaches to low energy quark dynamics, as e.g.\ the instanton
liquid model~\citep{SS98} and the Schwinger-Dyson resummation
techniques~\citep{Roberts:1994dr}. Lattice QCD
calculations~\citep{Parappilly:2005ei,Bowman2003,Furui:2006ks} also indicate
that quark interactions should act over a certain range in the momentum space.
Moreover, it has been argued that nonlocal extensions of the NJL model do not
show some of the above mentioned inconveniences of the local theory. Indeed,
nonlocal interactions regularize the model in such a way that anomalies are
preserved~\citep{RuizArriola:1998zi} and charges are properly quantized, the
effective interaction is finite to all orders in the loop expansion and
therefore there is not need to introduce extra
cutoffs~\citep{Blaschke:1995gr}, soft regulators such as Gaussian functions
lead to small next-to-leading order corrections~\citep{Rip00}, etc.

In the present work we will reconsider non-local models adopting as the basic
ingredient a reliable description of the quark propagator as given from
fundamental studies, such as lattice QCD. In this sense, it should be noticed
that, except for Ref.\citep{Noguera:2005ej,Noguera:2005cc}, most of the
calculation performed so far using non-local chiral quark models have
neglected the wave function renormalization in the propagator (See e.g. Refs.
\citep{Bowler:1994ir,Scarpettini:2003fj,GomezDumm:2006vz,Golli:1998rf,Rezaeian:2004nf}).
Recent lattice QCD calculations suggest, however, that such renomalization can
be of the order of 30 \% (or even more) at zero
momentum\citep{Parappilly:2005ei,Bowman2003,Furui:2006ks}. Moreover, these
calculations also show that the quark masses tend to their asymptotic values
in a rather soft way. Thus, it is of importance to perform a detailed study on
the incorporation of these features in this type of models, and analyze their
role in the prediction for different hadronic observables. The lagrangian we
will use is the minimal extension which allows to incorporate the full
momentum dependence of the quark propagator, through its mass and wave
function renormalization. Using this lagrangian we explore which are the
implications for some pion and sigma meson properties originated by changes in
the quark propagator. In particular, we present here results for the sigma
meson mass and width, and for the pion-pion scattering parameters. Studying
these scattering parameters close to the chiral limit we are also able to
obtain predictions for some of the low energy constants of the Chiral
Perturbation Theory ($\chi$PT) Lagrangian~\citep{Gasser:1983yg}.

The present article is organized as follows. In Sec. II we present the model
lagrangian and the formalism necessary to derive some selected pion and sigma
meson properties. In Sec. III we discuss different ways to obtain the model
parameters and compare the resulting quark propagators with available lattice
data. In Sec. IV we present and discuss the predictions of the model for the
selected parametrizations, paying special attention to the role played by the
incorporation of the wavefunction renormalization and by the difference in the
quark interaction momentum dependence. In Sec. V we analyze the connection of
the non-local quark models described here with $\chi$PT, and present the
predictions for the corresponding low energy constants relevant for $\pi$%
-$\pi$ scattering to $O(4)$ in the chiral expansion. Finally, in Sec. VI our
main conclusions are summarized.

\section{The model}

\subsection{Effective action}

Let us begin by stating the Euclidean action for the nonlocal chiral quark
model in the case of two light flavors,
\begin{equation}
S_{E}=\int d^{4}x\ \left\{  \bar{\psi}(x)\left(  -i\rlap/\partial
+m_{c}\right)  \psi(x)-\frac{G_{S}}{2}\Big[j_{a}(x)j_{a}(x)-j_{P}%
(x)j_{P}(x)\Big]\right\}  \ . \label{action}%
\end{equation}
Here $m_{c}$ is the current quark mass, which is assumed to be equal for $u$
and $d$ quarks. The nonlocal currents $j_{a}(x),j_{P}(x)$ are given by
\begin{align}
j_{a}(x)  &  =\int d^{4}z\ g(z)\ \bar{\psi}\left(  x+\frac{z}{2}\right)
\ \Gamma_{a}\ \psi\left(  x-\frac{z}{2}\right)  \ .\nonumber\\
j_{P}(x)  &  =\int d^{4}z\ f(z)\ \bar{\psi}\left(  x+\frac{z}{2}\right)
\ \frac{i{\overleftrightarrow{\rlap/\partial}}}{2\ \varkappa_{p}}\ \psi\left(
x-\frac{z}{2}\right)  \label{cuOGE}%
\end{align}
Here, $\Gamma_{a}=(\leavevmode\hbox{\small1\kern-3.8pt \normalsize1},i\gamma
_{5}\vec{\tau})$ and $u(x^{\prime}){\overleftrightarrow{\partial}%
}v(x)=u(x^{\prime})\partial_{x}v(x)-\partial_{x^{\prime}}u(x^{\prime})v(x)$.
The functions $g(z)$ and $f(z)$ in Eq.(\ref{cuOGE}), are nonlocal covariant
form factors characterizing the corresponding interactions. The four standard
quark currents, $j_{a}(x)$, require the same $g(z)$ form factor to guarantee
chiral invariance. The new term, $j_{P}(x)j_{P}(x),$ is self-invariant under
chiral transformations. The scalar-isoscalar component of the $j_{a}(x)$
current will generate the momentum dependent quark mass in the quark
propagator, while the \char`\"{}momentum\char`\"{} current, $j_{P}(x),$ will
be responsible for a momentum dependent wave function renormalization of this
propagator. For convenience, we take the same coupling parameter, $G_{S},$ for
the standard chiral quark interaction and for the new $j_{P}(x)j_{P}(x)$ term.
Note, however, that the relative strength between both interaction terms will
be controlled by the mass parameter $\varkappa_{p}$ introduced in
Eq.(\ref{cuOGE}). We have choosen the relative sign between these terms in
order to have a real value for $\varkappa_{p}$ for the case in which the wave
function renormalization $Z\left(  p\right)  $ (explicitly defined in
Eq.(\ref{propa} below) is less than 1. In what follows it is convenient to
Fourier transform $g(z)$ and $f(z)$ into momentum space. Note that Lorentz
invariance implies that the Fourier transforms $g(p)$ and $f(p)$ can only be
functions of $p^{2}$.

In order to deal with meson degrees of freedom, one can perform a standard
bosonization of the theory. This is done by considering the corresponding
partition function $\mathcal{Z}=\int\mathcal{D}\bar{\psi}\,\mathcal{D}%
\psi\,\exp[-S_{E}]$, and introducing auxiliary fields $\sigma_{1}%
(x),\sigma_{2}(x),\vec{\pi}(x))$, where $\sigma_{1,2}(x)$ and $\vec{\pi}(x)$
are scalar and pseudoscalar mesons, respectively. Integrating out the quark
fields we get
\begin{equation}
\mathcal{Z}=\int\mathcal{D}\sigma_{1}\,\mathcal{D}\sigma_{2}\ \mathcal{D}%
\vec{\pi}\ \exp[-S_{E}^{\mathrm{bos}}]\ ,
\end{equation}
where
\begin{equation}
S_{E}^{\mathrm{bos}}=-\ln\,\det A+\frac{1}{2G_{S}}\int\frac{d^{4}p}{(2\pi
)^{4}}\ \left[  \sigma_{1}(p)\ \sigma_{1}(-p)+\vec{\pi}(p)\cdot\vec{\pi
}(-p)+\sigma_{2}(p)\ \sigma_{2}(-p)\right]  .
\end{equation}
The operator $A$ reads, in momentum space,
\begin{align}
A(p,p^{\prime})\!  &  \!=\!\!(-\rlap/p+m_{c})\,(2\pi)^{4}\,\delta
^{(4)}(p-p^{\prime})+g\left(  \frac{p+p^{\prime}}{2}\right)  \ \left[
\sigma_{1}(p^{\prime}-p)+i\gamma_{5}\vec{\tau}\cdot\vec{\pi}(p^{\prime
}-p)\right] \nonumber\\
&  \qquad\qquad\qquad\qquad\qquad\qquad+\ f\left(  \frac{p+p^{\prime}}%
{2}\right)  \ \frac{\rlap/p+\rlap/p^{\prime}}{2\ \varkappa_{p}}\ \sigma
_{2}(p^{\prime}-p),
\end{align}

At this stage we assume that the $\sigma_{1,2}$ fields have nontrivial
translational invariant mean field values $\bar{\sigma}_{1,2}$, while the mean
field values of the pseudoscalar fields $\pi_{i}$ are zero. Thus we write
\begin{align}
\sigma_{1}(x)  &  =\bar{\sigma}_{1}+\delta\sigma_{1}(x)\\
\sigma_{2}(x)  &  =\varkappa_{p}\ \bar{\sigma}_{2}+\delta\sigma_{2}(x)\\
\vec{\pi}(x)  &  =\delta\vec{\pi}(x)
\end{align}
Replacing in the bosonized effective action, and expanding in powers of the
meson fluctuations, we get
\[
S_{E}^{\mathrm{bos}}\ =\ S_{E}^{\mathrm{MFA}}+S_{E}^{\mathrm{quad}}+\ ...
\]
Here the mean field action per unit volume reads
\begin{equation}
\frac{S_{E}^{\mathrm{MFA}}}{V^{(4)}}=-2N_{c}\int\frac{d^{4}p}{(2\pi)^{4}%
}\ \mbox{tr} \ln\left[  \mathcal{D}_{0}^{-1}(p)\right]  +\frac{\bar{\sigma
}_{1}^{2}}{2G_{S}}+\frac{\varkappa_{p}^{2}\ \bar{\sigma}_{2}^{2}}{2G_{S}}\ ,
\end{equation}
where the quark propagator in the mean field approximation $\mathcal{D}%
_{0}(p)$ is given by
\begin{equation}
\mathcal{D}_{0}(p)=\frac{Z(p)}{-\rlap/p+M(p)} \label{propa}%
\end{equation}
with
\begin{align}
Z(p)  &  =\left(  1-\bar{\sigma}_{2}\ f(p)\right)  ^{-1}\nonumber\\
M(p)  &  =Z(p)\left(  m_{c}+\bar{\sigma}_{1}\ g(p)\right)  \label{mz}%
\end{align}
The quadratic terms can be written as
\begin{equation}
S_{E}^{\mathrm{quad}}=\frac{1}{2}\int\frac{d^{4}p}{(2\pi)^{4}}\left[
G_{\sigma}(p^{2})\ \delta\sigma(p)\ \delta\sigma(-p)+G_{\sigma^{\prime}}%
(p^{2})\ \delta\sigma^{\prime}(p)\ \delta\sigma^{\prime}(-p)+\ G_{\pi}%
(p^{2})\ \delta\vec{\pi}(p)\cdot\delta\vec{\pi}(-p)\right]  \ , \label{quad}%
\end{equation}
where the $\sigma$ and $\sigma^{\prime}$ fields are related to $\sigma_{1}$
and $\sigma_{2}$ by
\begin{align}
\delta\sigma &  =\cos\theta\ \ \delta\sigma_{1}-\sin\theta\ \ \delta\sigma
_{2}\\
\delta\sigma^{\prime}  &  =\sin\theta\ \ \delta\sigma_{1}+\cos\theta
\ \ \delta\sigma_{2}\ ,
\end{align}
and the mixing angle $\theta$ is defined in such a way that there is no
$\sigma-\sigma^{\prime}$ mixing at the level of the quadratic action. The
function $G_{\pi}(p^{2})$ introduced in Eq.~(\ref{quad}) is given by
\begin{equation}
G_{\pi}(p^{2})=\frac{1}{G_{S}}-\,8\, N_{c}\int\frac{d^{4}q}{(2\pi)^{4}}%
\ g^{2}(q)\frac{Z(q^{+})Z(q^{-})}{D(q^{+})D(q^{-})}\left[  q^{+}\cdot
q^{-}+M(q^{+})M(q^{-})\right]
\end{equation}
with $q^{\pm}=q\pm p/2\,$ and $D(q)=q^{2}+M^{2}(q)$, while for the
$\sigma-\sigma^{\prime}$ system we have
\begin{equation}
G_{{\scriptsize \left(
\begin{array}
[c]{c}%
\sigma\\
\sigma^{\prime}%
\end{array}
\right)  }}(p^{2})=\frac{G_{\sigma_{1}\sigma_{1}}(p^{2})+G_{\sigma_{2}%
\sigma_{2}}(p^{2})}{2}\mp\sqrt{\left[  G_{\sigma_{1}\sigma_{2}}(p^{2})\right]
^{2}\!+\!\left[  \frac{G_{\sigma_{1}\sigma_{1}}(p^{2})-G_{\sigma_{2}\sigma
_{2}}(p^{2})}{2}\right]  ^{2}}%
\end{equation}
where
\begin{align}
G_{\sigma_{1}\sigma_{1}}(p^{2})  &  =\frac{1}{G_{S}}-\,8\, N_{c}\int
\frac{d^{4}q}{(2\pi)^{4}}\ g^{2}(q)\frac{Z(q^{+})Z(q^{-})}{D(q^{+})D(q^{-}%
)}\left[  q^{+}\cdot q^{-}-M(q^{+})M(q^{-})\right] \nonumber\\
G_{\sigma_{2}\sigma_{2}}(p^{2})  &  =\frac{1}{G_{S}}+\,\frac{8\, N_{c}%
}{\varkappa_{p}^{2}}\ \int\frac{d^{4}q}{(2\pi)^{4}}\ q^{2}f^{2}(q)\frac
{Z(q^{+})Z(q^{-})}{D(q^{+})D(q^{-})}\left[  (q^{+}\cdot q^{-})-M(q^{+}%
)M(q^{-})+\frac{(q^{+})^{2}(q^{-})^{2}-(q^{+}\cdot q^{-})^{2}}{2q^{2}}\right]
\nonumber\\
G_{\sigma_{1}\sigma_{2}}(p^{2})  &  =-\frac{8\, N_{c}}{\varkappa_{p}}%
\ \int\frac{d^{4}q}{(2\pi)^{4}}\ g(q)f(q)\frac{Z(q^{+})Z(q^{-})}%
{D(q^{+})D(q^{-})}\ q\cdot\left[  q^{-}M(q^{+})+q^{+}M(q^{-})\right]
\end{align}

\subsection{Mean field approximation and chiral condensates}

In order to find the mean field values $\bar{\sigma}_{1,2}$, one has to
minimize the action $S_{E}^{\mathrm{MFA}}$. A straightforward exercise leads
to the coupled gap equations
\begin{align}
\bar{\sigma}_{1}-8N_{c}\ G_{S}\int\frac{d^{4}p}{(2\pi)^{4}}\ g(p)\ \frac
{Z(p)M(p)}{D(p)}  &  =0\nonumber\\
\bar{\sigma}_{2}+8N_{c}\ G_{S}\int\frac{d^{4}p}{(2\pi)^{4}}\ \frac{p^{2}%
}{\varkappa_{p}^{2}}\ f(p)\ \frac{Z(p)}{D(p)}  &  =0. \label{gapeq}%
\end{align}

Now the chiral condensates are given by the vacuum expectation values
$\langle\bar{q}q\rangle=\langle\bar{u}u\rangle=\langle\bar{d}d\rangle$. They
can be easily obtained by performing the variation of $\mathcal{Z}%
^{\mathrm{MFA}}=\exp[-S_{E}^{MFA}]$ with respect to the corresponding current
quark masses. This expression turns out to be divergent. Thus, as customary,
we regularize it by subtracting its value for non-interacting quarks. We
obtain
\[
\langle\,\bar{q}\, q\,\rangle=-\,4N_{c}\int\frac{d^{4}p}{(2\pi)^{4}}\ \left(
\frac{Z(p)M(p)}{D(p)}-\frac{m_{c}}{p^{2}+m_{c}^{2}}\right)  .
\]

\subsection{Meson masses and quark-meson coupling constants}

The meson masses can be obtained by solving the equation
\begin{align}
G_{M}(-m_{M}^{2})=0\ . \label{pieq}%
\end{align}

In the case of the $\sigma-\sigma^{\prime}$ system the mixing angles is given
by $\theta(-m_{\sigma,\sigma^{\prime}}^{2}),$ where
\begin{equation}
\tan2\ \theta(p^{2})=\frac{2G_{\sigma_{1}\sigma_{2}}(p^{2})}{G_{\sigma
_{2}\sigma_{2}}(p^{2})-G_{\sigma_{1}\sigma_{1}}(p^{2})}%
\end{equation}

Finally, the on-shell meson-quark coupling constants $g_{Mq\bar{q}}$ are given
by
\begin{equation}
g_{Mq\bar{q}}^{-2}\equiv G_{Mq\bar{q}}^{-2}(-m_{M}^{2})\ =\ \frac{dG_{M}%
(p)}{dp^{2}}\bigg|_{p^{2}=-m_{M}^{2}}\ . \label{gpiqq}%
\end{equation}

Note that due to the mixing, in the scalar meson channel the corresponding
vertex has two components. Thus for $\sigma q\bar{q}$ vertex we have
\begin{equation}
V_{\sigma q\bar{q}}=g_{\sigma q\bar{q}}^{0}%
\ \leavevmode\hbox{\small1\kern-3.8pt \normalsize1}+g_{\sigma q\bar{q}}%
^{1}\frac{\rlap/p+\rlap/p^{\prime}}{2\varkappa_{p}}%
\end{equation}
where
\begin{equation}
g_{\sigma q\bar{q}}^{(0)}=g_{\sigma q\bar{q}}\ \cos\theta\qquad;\qquad
g_{\sigma q\bar{q}}^{(1)}=g_{\sigma q\bar{q}}\ \sin\theta
\end{equation}

\subsection{Pion weak decay constant}

By definition the pion weak decay constant $f_{\pi}$ is given by the matrix
element of the axial current $A_{\mu}^{a}(x)$ between the vacuum and the
renormalized one-pion state at the pion pole:
\begin{equation}
\langle0|A_{\mu}^{a}(0)|\tilde{\pi}^{b}(p)\rangle=i\ \delta^{ab}\ p_{\mu
}\ f_{\pi}\ .
\end{equation}

In order to obtain an explicit expression for the axial current, we have to
{}``gauge'' the effective action $S_{E}$ by introducing a set of axial gauge
fields $\mathcal{A}_{\mu}^{a}(x)$. For a local theory this {}``gauging''
procedure is usually done by performing the replacement
\begin{equation}
\partial_{\mu}\rightarrow\partial_{\mu}+\frac{i}{2}\ \gamma_{5}\ \vec{\tau
}\cdot{\vec{\mathcal{A}}}_{\mu}(x)\ .
\end{equation}
In the present case ---owing to the nonlocality of the involved fields--- one
has to perform additional replacements in the interaction terms. Namely,
\begin{align}
\psi(x-z/2)\ \rightarrow\ W_{A}\left(  x,x-z/2\right)  \ \psi
(x-z/2)\nonumber\\
\psi^{\dagger}(x+z/2)\ \rightarrow\ \psi^{\dagger}(x+z/2)\ W_{A}\left(
x+z/2,x\right)  \label{gauge}%
\end{align}
Here $x$ and $z$ are the variables appearing in the definitions of the
nonlocal currents (see Eq.(\ref{cuOGE})), and the function $W_{A}(x,y)$ is
defined by
\begin{equation}
W_{A}(x,y)\ =\ \mathrm{P}\;\exp\left[  \frac{i}{2}\int_{x}^{y}ds_{\mu}%
\ \gamma_{5}\ \vec{\tau}\cdot{\vec{\mathcal{A}}}_{\mu}(s)\right]  \ ,
\end{equation}
where $s$ runs over an arbitrary path connecting $x$ with $y$.

Once the gauged effective action is built, it is easy to get the axial current
as the derivative of this action with respect to $\mathcal{A}_{\mu}^{a}(x)$,
evaluated at ${\vec{\mathcal{A}}}_{\mu}(x)=0$. Performing the derivative of
the resulting expressions with respect to the renormalized meson fields, we
can finally identify the corresponding meson weak decay constants. After a
rather lengthy calculation we obtain
\begin{equation}
f_{\pi}=\frac{m_{c}\; g_{\pi q\bar{q}}}{m_{\pi}^{2}}\; F_{0}(-m_{\pi}^{2})\ .
\label{relation}%
\end{equation}
with
\begin{equation}
F_{0}(p^{2})=8\, N_{c}\int\frac{d^{4}q}{(2\pi)^{4}}\ g(q)\;\frac
{Z(q^{+})Z(q^{-})}{D(q^{+})D(q^{-})}\ \left[  q^{+}\cdot q^{-}+M(q^{+}%
)M(q^{-})\right]
\end{equation}
It is important to notice that the integration over the path variable $s$
appearing in this calculation turns out to be trivial and, thus, the result
path-independent. In the chiral limit the expression Eq.(\ref{relation}) has a
rather simple form \citep{Noguera:2005ej} given by
\begin{equation}
f_{\pi}=\frac{M\left(  0\right)  }{g_{\pi qq}Z\left(  0\right)  }\ \ \ \ ,
\end{equation}
which connects with the Goldberger-Treiman relation.

\subsection{The decay width of the Sigma meson}

To obtain the decay amplitude of the $\sigma$ meson into two pion we need to
calculate
\begin{equation}
\frac{\delta S_{E}^{\mathrm{bos}}}{\delta\sigma(q)\delta\pi^{a}(q_{1}%
)\delta\pi^{b}(q_{2})}=(2\pi)^{4}\ \delta^{4}(q+q_{1}+q_{2})\ \delta
_{ab}\ G_{\sigma\pi\pi}(q^{2},q_{1}^{2},q_{2}^{2})
\end{equation}
where the meson fields are assumed to be already renormalized. In terms of the
unrenormalized fields and taking into account the $\sigma_{1}-\sigma_{2}$
mixing we have
\begin{equation}
G_{\sigma\pi\pi}(q^{2},q_{1}^{2},q_{2}^{2})=G_{\sigma q\bar{q}}(q^{2})\ G_{\pi
q\bar{q}}(q_{1}^{2})\ G_{\pi q\bar{q}}(q_{2}^{2})\ \tilde{G}_{\sigma\pi\pi
}(q^{2},q_{1}^{2},q_{2}^{2})
\end{equation}
where
\begin{equation}
\tilde{G}_{\sigma\pi\pi}(q^{2},q_{1}^{2},q_{2}^{2})=G_{\sigma_{1}\pi\pi}%
(q^{2},q_{1}^{2},q_{2}^{2})\ \cos\theta(q^{2})-G_{\sigma_{2}\pi\pi}%
(q^{2},q_{1}^{2},q_{2}^{2})\ \sin\theta(q^{2})
\end{equation}
and the expressions of the unrenormalized $\sigma_{1}$ and $\sigma_{2}$
coupling constants to two $\pi$ can be obtained by expanding $\Gamma$ to third
order in the fluctuations. We get
\begin{align}
G_{\sigma_{1}\pi\pi}(q^{2},q_{1}^{2},q_{2}^{2})  &  =-16N_{c}\int\frac{d^{4}%
k}{(2\pi)^{4}}\ \ g\left(  \frac{k_{1}+k_{2}}{2}\right)  g\left(
\frac{k+k_{1}}{2}\right)  g\left(  \frac{k+k_{2}}{2}\right)  \frac
{Z(k)Z(k_{1})Z(k_{2})}{D(k)D(k_{1})D(k_{2})}\times\nonumber\\
&  \qquad\qquad\times\Big[M(k)\ k_{1}\cdot k_{2}-M(k_{1})\ k\cdot
k_{2}-M(k_{2})\ k\cdot k_{1}-M(k)\ M(k_{1})\ M(k_{2})\Big]\\
\nonumber\\
G_{\sigma_{2}\pi\pi}(q^{2},q_{1}^{2},q_{2}^{2})  &  =-\frac{8N_{c}}%
{\varkappa_{p}}\ \int\frac{d^{4}k}{(2\pi)^{4}}\ \ f\left(  \frac{k_{1}+k_{2}%
}{2}\right)  g\left(  \frac{k+k_{1}}{2}\right)  g\left(  \frac{k+k_{2}}%
{2}\right)  \frac{Z(k)Z(k_{1})Z(k_{2})}{D(k)D(k_{1})D(k_{2})}\times\nonumber\\
&  \!\!\!\!\!\!\!\!\!\!\!\!\!\!\!\times\Big[k_{2}^{2}\ k\cdot k_{1}+k_{1}%
^{2}\ k\cdot k_{2}+\left(  k_{1}+k_{2}\right)  \cdot\left[  M(k_{1}%
)\ M(k)\ k_{2}-M(k_{2})\ M(k_{1})\ k+M(k_{2})\ M(k)\ k_{1}\right]  \Big]
\end{align}
where $k_{1}=k+q_{1}$ and $k_{2}=k-q_{2}$ and $q^{2}=(q_{1}+q_{2})^{2}$.
Similarly for $\sigma^{\prime}$ we have
\begin{equation}
G_{\sigma^{\prime}\pi\pi}(q^{2},q_{1}^{2},q_{2}^{2})=G_{\sigma^{\prime}%
q\bar{q}}(q^{2})\ G_{\pi q\bar{q}}(q_{1}^{2})\ G_{\pi q\bar{q}}(q_{2}%
^{2})\ \tilde{G}_{\sigma^{\prime}\pi\pi}(q^{2},q_{1}^{2},q_{2}^{2})
\end{equation}
where
\begin{equation}
\tilde{G}_{\sigma^{\prime}\pi\pi}(q^{2},q_{1}^{2},q_{2}^{2})=G_{\sigma_{1}%
\pi\pi}(q^{2},q_{1}^{2},q_{2}^{2})\ \sin\theta(q^{2})+G_{\sigma_{2}\pi\pi
}(q^{2},q_{1}^{2},q_{2}^{2})\ \cos\theta(q^{2})
\end{equation}

In terms of $g_{M\pi\pi}=G_{M\pi\pi}(m_{M}^{2},m_{\pi}^{2},m_{\pi}^{2})$ the
$M=\sigma,\sigma^{\prime}$ width reads
\begin{equation}
\Gamma_{M\rightarrow\pi\pi}=\frac{3}{2}\frac{g_{M\pi\pi}^{2}}{16\pi m_{M}%
}\sqrt{1-\frac{4m_{\pi}^{2}}{m_{M}^{2}}}%
\end{equation}

\subsection{$\pi$-$\pi$ scattering}

In general, the total amplitude for the $\pi$-$\pi$ scattering process can be
expressed as
\begin{equation}
\mathcal{A}\left(  \pi^{\alpha}(q_{1})+\pi^{\beta}(q_{2})\rightarrow
\pi^{\gamma}(q_{3})+\pi^{\delta}(q_{4})\right)  =\delta_{\alpha\beta}%
\delta_{\gamma\delta}A(s,t,u)+\delta_{\alpha\gamma}\delta_{\beta\delta
}A(t,s,u)+\delta_{\alpha\delta}\delta_{\beta\gamma}A(u,t,s)
\end{equation}
where
\begin{equation}
s=(q_{1}+q_{2})^{2}\qquad;\qquad t=(q_{1}-q_{3})^{2}\qquad;\qquad
u=(q_{1}-q_{4})^{2} \label{mand}%
\end{equation}
Within the present model, this amplitude gets two contributions. One
corresponds to the box diagram and the other to the scalar meson pole diagram.
Thus,
\begin{equation}
A(s,t,u)=A_{box}(s,t,u)-g_{\pi q\bar{q}}^{4}\sum_{M=\sigma,\sigma^{\prime}%
}\tilde{G}_{M\pi\pi}^{2}(s,m_{\pi}^{2},m_{\pi}^{2})\ G_{M}^{-1}(s)
\end{equation}
where
\begin{equation}
A_{box}(s,t,u)=g_{\pi q\bar{q}}^{4}\ \left[
J(s,t,u)+J(s,u,t)-J(u,t,s)\right]
\end{equation}
and
\begin{equation}
J\left(  s,t,u\right)  =\frac{1}{2}\left[  J_{box}\left(  q_{1},q_{2}%
,q_{3}\right)  +J_{box}\left(  q_{1},-q_{3},-q_{2}\right)  \right]
\label{jota}%
\end{equation}
with%
\begin{align}
J_{box}\left(  q_{1},q_{2},q_{3}\right)   &  =16N_{c}\int\frac{d^{4}k}{2\pi
}\ g\left(  \frac{k+k_{1}}{2}\right)  \ g\left(  \frac{k+k_{2}}{2}\right)
\ g\left(  \frac{k_{1}+k_{13}}{2}\right)  \ g\left(  \frac{k_{2}+k_{13}}%
{2}\right)  \frac{Z(k_{1})Z(k)Z(k_{2})Z(k_{13})}{D(k_{1})D(k)D(k_{2}%
)D(k_{13})}\times\nonumber\\
&  \Big\{\left[  k_{1}\cdot k+M(k_{1})M(k)\right]  \left[  k_{2}\cdot
k_{13}+M(k_{1})M(k_{13})\right]  -\left[  k_{1}\cdot k_{2}+M(k_{1}%
)M(k_{2})\right]  \left[  k\cdot k_{13}+M(k)M(k_{13})\right] \nonumber\\
&  +\left[  k_{1}\cdot k_{13}+M(k_{1})M(k_{13})\right]  \left[  k\cdot
k_{2}+M(k)M(k_{2})\right]  \Big\} \label{jotaBox}%
\end{align}
where $k_{1}=k+q_{1}$, $k_{2}=k-q_{2}$, $k_{13}=k+q_{1}-q_{3}.$

It is customary to define the scattering amplitudes of defined isospin
\begin{align}
T^{0}  &  =3A(s,t,u)+A(t,s,u)+A(u,t,s)\nonumber\\
T^{1}  &  =A(t,s,u)-A(u,t,s)\nonumber\\
T^{2}  &  =A(t,s,u)+A(u,t,s)
\end{align}

In terms of these amplitudes the scattering lengths $a_{\ell}^{I}$ and slope
parameters $b_{\ell}^{I}$ are defined by the partial wave expansion at low
$q^{2}$
\begin{align}
\frac{1}{64\pi m_{\pi}}\int_{-1}^{1}dx\ P_{\ell}(x)\ T^{I}(s,t,u)=q^{2\ell
}\left(  a_{\ell}^{I}+b_{\ell}^{I}\ q^{2}+...\right)  \label{Texp}%
\end{align}
where $P_{\ell}(x)$ is the Lagrange polynomial of order $l$.

\section{Determination of the model parameters}

In this section we present in some detail the procedure used to determine the
model parameters as well as the form factors $g(q)$ and $f(q)$ which
characterize the non-local interactions.

In our first model (scenario S1) we use exponential functions to model the
non-local interactions. These are well behaved functions which have been often
used in the literature (see e.g.
\citep{Bowler:1994ir,Scarpettini:2003fj,GomezDumm:2006vz,Golli:1998rf}) to
define $g(q)$. Here, we also use such form for $f(q)$. Thus, for S1 we have
\begin{equation}
g(p)=\exp(-p^{2}/\Lambda_{0}^{2})\qquad;\qquad f(p)=\exp(-p^{2}/\Lambda
_{1}^{2})
\end{equation}
Note that the range (in momentum space) of the nonlocality in each channel is
determined for the parameters $\Lambda_{0}$ and $\Lambda_{1}$, respectively.
From Eq. (\ref{mz}) we obtain
\begin{align}
M\left(  p\right)   &  =Z\left(  p\right)  \ \left[  m_{c}+\bar{\sigma}%
_{1}\ \exp(-p^{2}/\Lambda_{0}^{2})\right] \nonumber\\
Z\left(  p\right)   &  =\left[  1-\bar{\sigma}_{2}\ \exp(-p^{2}/\Lambda
_{1}^{2})\right]  ^{-1}%
\end{align}
We fix the values of $m_{c}$ and $<q\bar{q}>^{1/3}$ to reasonable values
$m_{c}=5.7$ MeV and $<q\bar{q}>^{1/3}=-240$ MeV determining the rest of the
parameters so as to reproduce the empirical values $f_{\pi}=92.4$ MeV and
$m_{\pi}=139$ MeV, and $Z(0)=0.7$ which is within the range of values
suggested by recent lattice calculations\citep{Parappilly:2005ei,Furui:2006ks}.

For the second parametrization we follow Ref.\citep{Noguera:2005ej}, where a
parametrization based on a fit to the mass and renormalization functions
obtained in a Landau gauge lattice calculation was used. Such parametrization
is
\begin{align}
M(p)  &  =m_{c}+\alpha_{m}\ f_{m}(p)\ \ ,\nonumber\\
Z(p)  &  =1+\alpha_{z}\ f_{z}(p)\ \ \ \ \ ,
\end{align}
with%
\begin{equation}
f_{m}(p)=\left[  1+\left(  p^{2}/\Lambda_{0}^{2}\right)  ^{3/2}\right]
^{-1}\ \qquad;\qquad f_{z}(p)=\left[  1+\left(  p^{2}/\Lambda_{1}^{2}\right)
\right]  ^{-5/2}\ \ \ ,
\end{equation}
where the analytical form of $f_{m}\left(  p\right)  $ has been proposed in
Ref.\citep{Bowman2003}. The analytical form of $f_{z}\left(  p\right)  $ is
chosen in order to guarantee the convergence of the integrals. Some
alternative parametrization of this type suggested from vector meson dominance
of the pion form factor can be found in Ref.\citep{RuizArriola:2003bs}. In
terms of the functions $f_{m}(p)$ and $f_{z}(p)$, and the constants
$m_{c},\alpha_{m},\alpha_{z}$ the form factors $g(q)$ and $f(q)$ are given by
\begin{align}
g(p)  &  =\frac{1+\alpha_{z}}{1+\alpha_{z}f_{z}(p)}\frac{\alpha_{m}%
f_{m}(p)-m_{c}\alpha_{z}f_{z}(p)}{\alpha_{m}-m_{c}\alpha_{z}}%
\ \ \ ,\nonumber\\
f(p)  &  =\frac{1+\alpha_{z}}{1+\alpha_{z}f_{z}(p)}\ f_{z}(p)\ \ \ \ .
\end{align}
and the mean field values are
\begin{align}
\bar{\sigma}_{1}  &  =\frac{\alpha_{m}-m_{c}\alpha_{z}}{1+\alpha_{z}%
}\nonumber\\
\bar{\sigma}_{2}  &  =\frac{\alpha_{z}}{1+\alpha_{z}}%
\end{align}
The parameters for this second model (scenario S2) are determine as follows.
As before we take $Z(0)=0.7$ and fix $\Lambda_{0}$ and $\Lambda_{1}$ in such a
way that the functions $f_{m}\left(  p\right)  $ and $Z\left(  p\right)  $
agree reasonable well with lattice results of Ref.\citep{Parappilly:2005ei}.
Next we fix $m_{c}$ and $\alpha_{m}$ in order to reproduce the physical values
of $m_{\pi}$ and $f_{\pi}.$ The resulting parameters are $m_{c}=2.37$ MeV,
$\alpha_{m}=309$ MeV, and with $\Lambda_{0}=850$ MeV and $\Lambda_{1}=1400$ MeV.

Finally, in order to compare with previous studies where the wavefunction
renormalization of the quark propagator has been ignored we consider a third
model (scenario S3). In such scenario we take $Z(p)=1$ and exponential
parametrization for $g(p)$. Such model corresponds to the \char`\"{}Model
II\char`\"{} discussed in Ref.\citep{GomezDumm:2006vz}, from where we take the
parameters corresponding to $<q\bar{q}>^{1/3}=-240$ MeV.

\begin{figure}[h]
\includegraphics[width=0.8\textwidth]{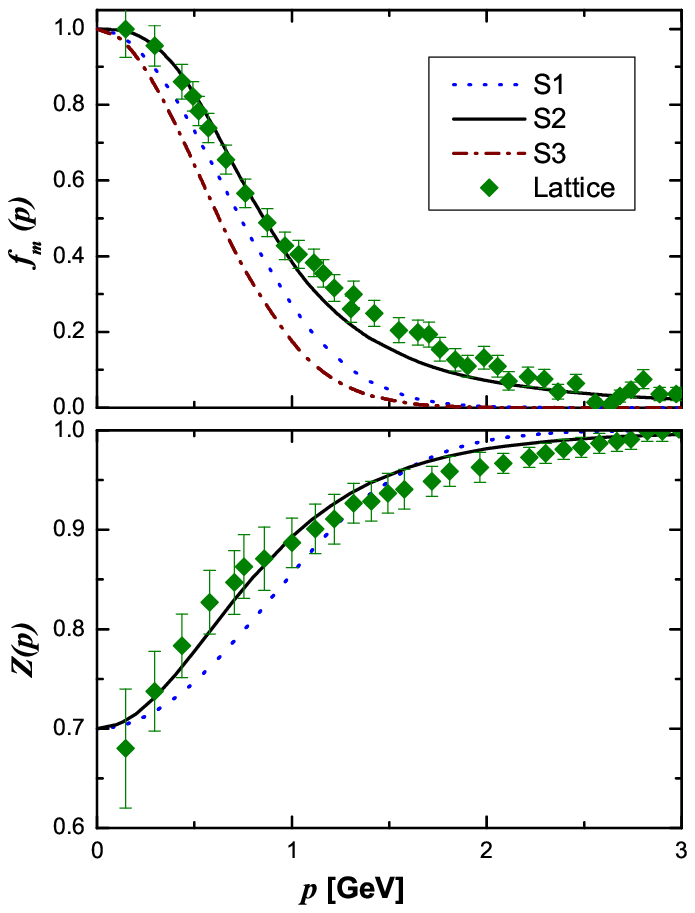}\caption{(Color online) $f_{m}(p)$
(Upper panel) and $Z(p)$ (lower panel) for various parametrization as compared
with Lattice results of Ref. \citep{Parappilly:2005ei}}%
\label{Fig1}%
\end{figure}

The values of the model parameters for each of the chosen scenarios are
summarized in Table \ref{tab1}. In Fig.\ref{Fig1} we compare the quark mass
function $f_{m}(p)$ and renormalization function $Z(p)$ as obtained from our
three scenarios with data extracted from the lattice results of
Ref.\citep{Parappilly:2005ei}. The main reason for comparing $f_{m}(p)$
(instead of $M\left(  p\right)  $) is that analyzing lattice data from
different groups using Landau gauge
fixing\citep{Parappilly:2005ei,Furui:2006ks}, and also results for $M\left(
p\right)  $ obtained by each group using different inputs, we observed that
the resulting functions $f_{m}(p)$ are very similar in spite of the
differently looking $M(p)$. On the other hand, the renomalization functions
$Z\left(  p\right)  $ are much less sensitive to the choice of lattice
parameters, and in fact the two lattice groups
\citep{Parappilly:2005ei,Furui:2006ks} provide similar results. We observe
that the functions $f_{m}\left(  p\right)  $ and $Z\left(  p\right)  $ for
scenario S1, based on exponential functions, decrease faster than the lattice
data. For scenario S2, however, they go to zero as $(p^{2})^{-3/2}$ and
$(p^{2})^{-5/2},$ respectively, following the lattice data in a closer manner.
Finally, in the case of S3 the exponential decrease of $f_{m}(p)$ is even
faster than that of $S_{1}$.

\section{Numerical results}

In this section we present and discuss our numerical results. In Table
\ref{tab1} we give the results for the mean-field properties, together with
the pion and sigma masses and decay parameters. As it can be seen in this
table, while for the exponential parameterizations (i.e. S1 and S3) the
empirical values of $f_{\pi}$ and $m_{\pi}$ are consistent with a quark
condensate which lies within the range of the usually quoted phenomenological
values $-\langle\bar{q}q\rangle^{1/3}\,\simeq\,200$ - $260$%
~MeV~\citep{Dosch:1997wb,Giusti:1998wy} the scenario S2 leads to a value of
the chiral condensate somewhat above such range. On the other hand, the
corresponding current quark mass is quite smaller than those obtained for the
scenarios S1 and S3. This issue deserves some comment. The chiral condensate,
as well as the current quark masses, are scale dependent objects. In
particular, the phenomenological values quoted above for the condensate
correspond to a choice of the renormalization scale $\mu=1$~GeV. In the
parametrization S2 some parameters have been determined so as to obtain a good
approximation to the lattice mass renormalization function $Z(p)$, a quantity
which also depends on the renormalization point. In particular, we use the
function $Z(p)$ obtained in Ref.\citep{Parappilly:2005ei} where the
renormalization scale has been chosen to be $\mu=3$ GeV. One might wonder
whether the fact that this renormalization point differs from the one usually
used to quote the values of the condensate can account for the fact that the
S2 prediction is outside the empirical range. If one assumes that this
difference is also responsible for the rather low value of $m_{c}$ this can be
investigated in the following way. To leading order in the chiral expansion
the current quark mass and the condensate are related by the
Gell-Mann-Oakes-Renner (GMOR) relation
\begin{equation}
f_{\pi}^{2}\ m_{\pi}^{2}=2<\bar{q}q>\hat{m}%
\end{equation}
where $\hat{m}=\left(  m_{u}+m_{d}\right)  /2$. The validity of GMOR to that
order is well justified by the low energy behavior of the $\pi\pi$ scattering
amplitudes \citep{Colangelo:2001sp}. Using that, according to
Ref.\citep{Yao:2006px}, $\hat{m}$ runs from ~ 5.5 MeV at the scale $\mu=1$ GeV
to 4.1 MeV at $\mu=2$ GeV we expect that a typical value of $\langle\bar
{q}q\rangle^{1/3}=-240$~MeV at $\mu=1$ GeV will run to $\langle\bar{q}%
q\rangle^{1/3}=-270$~MeV at $\mu=2$ GeV \citep{Jamin:2002ev}. Lattice
calculations provide an independent determination of quark masses and $\bar
{q}q$ condensate \citep{Gimenez:2005nt,Gimenez:2005va,Becirevic:2005qk}:
\begin{align}
m_{ud}^{\overline{MS}}\left(  2\ GeV\right)   &  =4.3\pm0.4_{stat}%
\genfrac{.}{.}{0pt}{}{+1.1}{-0.4}%
_{sys}MeV\label{Mas_lat}\\
\left\langle \bar{q}q\right\rangle \left(  2\ GeV\right)   &  =-\left(
265\pm5_{stat}\pm22_{sys}MeV\right)  ^{3}%
\end{align}
which confirms the $\mu=2$ GeV values given above. Note that since these two
lattice calculations are not connected, the quoted values imply a verification
of the GMOR relation. Since the GMOR relation is well satisfied by our
lagrangian model \citep{Noguera:2005ej}, and in all our scenarios $f_{\pi}$
and $m_{\pi}$ are fitted to their empirical values, it is clear that the
quality of the description of the quark condensate and the current quark mass
are closely related. Thus, a further running up to $\mu=3\ GeV$ implies that
the current quark mass must be scaled by a factor of the $\hat{m}%
(2\ \mbox{GeV})/\hat{m}(3\ \mbox{GeV})=1.11$. This value is rather different
from the factor 1.81 obtained from the ratio between the lattice result at
$\mu=2$ GeV and the value of $m_{c}$ for the scenario S2 given in Table
\ref{tab1}. This clearly indicates that possible ambiguities related to the
choice of renormalization point cannot fully account for the rather high value
of the condensate for the scenario S2. In fact, using the above mentioned
factors to reescale the value $-\langle\bar{q}q\rangle^{1/3}\,\simeq\,326$
~MeV quoted in Table \ref{tab1} down to $\mu=1\ GeV$ we get $-\langle\bar
{q}q\rangle^{1/3}\,\simeq\,284$ ~MeV which is about $10\%$ above the empirical
upper limit. A possible way to reduce the value of the quark condensate in S2
is to reduce the parameter $\Lambda_{0}$. For $\Lambda_{0}\sim600$ MeV we can
obtain values for the quark condensate and quark masses which are within the
phenomenological bounds.

The mass and width of the sigma meson display some dependence on the
parametrization. However, such dependence is smaller than the one found in the
local NJL model\citep{Nakayama:1991ue}. The obtained values for the masses are
somewhat larger than the recently extracted empirical values $478_{-23}%
^{+24}\pm17MeV$ \citep{Aitala:2000xu} and $390_{-36}^{+60}MeV$
\citep{Wu:2001vz} while the widths are compatible with the experimentally
reported values $324_{-40}^{+42}\pm21MeV$ \citep{Aitala:2000xu} and
$282_{-50}^{+77}MeV$ \citep{Wu:2001vz}.

The situation concerning the $\sigma^{\prime}$ meson deserves some comment. In
general, for the non-local models under consideration the quark propagators
develop a series of poles in the complex plane. In Euclidean space, such poles
can be purely imaginary (as in the NJL model which only has one pole of this
type) or fully complex. The existence of these poles implies the appearance of
\char`\"{}pinch points\char`\"{} \citep{Bowler:1994ir} in the calculation of
the meson two-point functions. The external momentum for which the first of
such \char`\"{}pinch points\char`\"{} appears is given by $p_{pp}=2S_{i}$
where $S_{i}$ is the imaginary part of the first pole of the quark propagator.
From this point on the functions $G$ in Eq.(17) do in general develop an
imaginary component related to the unphysical decay into $q\bar{q}$ pairs,
which is usually associated with the lack of confinement. In some cases,
depending on the regulator and/or parametrization, one can find a prescription
for the integration path along the complex plane such that this imaginary
component cancels
out\citep{Bowler:1994ir,Scarpettini:2003fj,Praszalowicz:2001wy}. It is clear,
however, that the corresponding results turn out to be prescription dependent
and, unless the meson pole appears no far above $p_{pp}$, not very reliable.
For this reason, in this work we take the point of view that $p_{pp}$ marks
the limit of validity of our model. For the three scenarios under
consideration we have found $p_{pp}$ to be about 1 GeV, which appears to be a
reasonable scale for a low energy effective model of QCD. As for the
$\sigma^{\prime}$ channel we have verified that no pole corresponding to a
meson of this type appears below that scale.

We turn now to the low-energy parameters for $\pi-\pi$ scattering. These
parameters have been matter of much attention in the recent past years. In
particular, recent results on Kl4 decays \citep{Pislak:2001bf,Aloisio:2002bs}
have led to an improved phenomenological
determination\citep{Colangelo:2001df,Kaminski:2006qe} of the threshold
parameters for S-, P-, D- and F- waves. Our results for the S and P waves are
displayed in Table \ref{tab2} while those corresponding to D and F waves in
Table \ref{tab3}. Since the calculation of sigma pole contributions include
off-shell quantities it is not possible to perform a clear and unique
separation between $\sigma$ and $\sigma^{\prime}$ contributions. Thus, only
the sum of such contributions is given. In general, reasonable estimates
indicate that $\sigma^{\prime}$ contributions represent only a few percent of
this total value. The phenomenological values extracted in
Ref.\citep{Kaminski:2006qe} are also indicated. In comparing our results with
these values one should keep in mind that the present model does not
incorporate pion loops, and hence there is still room for improvement.
Finally, for comparison, in Tables \ref{tab2} and \ref{tab3} the existing
results for the local SU(2) NJL model \citep{Schuren:1991sc,Bernard:1992mp}
are given. Results obtained in alternative QCD-based quark models can be
found, e.g. in Ref.\citep{sdyson}

We analyze first the results corresponding to the S- and P-waves. Let us
recall that to leading order in the chiral expansion the corresponding length
and slope parameters can be obtained from the Weinberg amplitude
\begin{equation}
A\left(  s,t,u\right)  =\frac{s-f_{\pi}^{2}}{m_{\pi}^{2}}%
\end{equation}
which leads to the predictions
\begin{equation}
\frac{8}{7}\ a_{0}^{0}\cdot m_{\pi}=-4\ a_{0}^{2}\cdot m_{\pi}=b_{0}^{0} \cdot
m_{\pi}^{3}=-2\ b_{0}^{2}\cdot m_{\pi}^{3}=6\ a_{1}^{1}\cdot m_{\pi}^{3}
=\frac{m_{\pi}^{2}}{4~\pi~f_{\pi}^{2}} \label{ScatLengthChiral}%
\end{equation}
Since our three different scenarios lead to the same values of $f_{\pi}$ and
$m_{\pi}$ the predictions for these five scattering parameters are expected to
be quite similar. In fact, results in Table \ref{tab2} confirm this, although
those of S2 are in slightly better agreement with empirical data. This is
particularly interesting in the case of $a_{0}^{2}$, which results from a
rather strong cancellation between box and sigma contributions. In order to be
more sensitive in the comparison between scenarios, we also give in Table
\ref{tab2} the combination of the S-wave isospin 0 and 2 parameters
$2a_{0}^{0}+7a_{0}^{2}$ which vanishes in the chiral limit. We observe that in
all scenarios the correction goes in the right direction. Moreover, in the
case of S2 its magnitude is larger providing therefore a better description of
the experimental result. Another way to improve on the discrimination between
the different parametrizations of our model is to consider corrections up to
$q^{6}$ order in the expansion Eq.(\ref{Texp}). Thus, we calculate the
parameters $c_{l}^{I}$ and $d_{l}^{I}$ corresponding to the $q^{4}$ and
$q^{6}$ corrections, respectively. We observe that in each partial wave the
exponential interaction produces scattering parameters which decrease rather
fast with the power of $q^{2}$. On the other hand, the scenario S2 predicts
coefficients which are of the same order of magnitude in each partial wave.

We consider now the scattering lengths and slope parameters for D- and F-waves
displayed in Table \ref{tab3}. These results, together with the scattering
lengths and slope parameters of S- and P-waves given in Table \ref{tab2},
complete all cases for which there are phenomenological determinations
available. For S1 and S3, we observe that the signs of the parameters are
correctly predicted, except for $b_{2}^{0}$ in S3. The absolute values for the
scattering lengths are off by a factor between 1.5 and 2.5, whereas the slope
parameters fail by one order of magnitude. On the other hand, the scenario S2
gives the right sign and order of magnitude in all cases, deviating only by a
factor 3 in the worse case, $b_{3}^{1}$.

From the previous results we can conclude that although the exponential
interaction might be able to reproduce the scattering lengths parameters
rather well the description of higher power coefficients is, in general,
expected to be less accurate as the power in $q^{2}$ increases. This is
particularly so for the higher partial waves. On the other hand, the momentum
dependence of the scenario S2 seems to be better adapted for the description
of the higher power parameters. In fact, the only case where this scenario
gives a worse result than the exponential ones is in the prediction for
$b_{1}^{1}$, where a strong cancellation between the box and sigma
contribution takes place.

Comparing scenarios S1 and S3 we can observe the effect on the scattering
parameters of taking into account the wave function renormalization. Except
for the parameters listed in Eq.(\ref{ScatLengthChiral}), we observe that as
the power in $q^{2}$ increases the associated parameters obtained in scenario
S3 decrease faster than in scenario S1. We can conclude that the effect of the
wave function renormalization term goes in the right direction, even if this
effect is less important than the one produced by the difference in the
momentum dependence of the interactions. It should be noticed that our
scenario S3 is very similar to the model used in Ref.\citep{Osipov:2006js}. In
fact, in both cases the wave function renormalization is not included,
exponential parameterizations are used and the values of $m_{\pi}$ and
$f_{\pi}$ are fitted. The difference comes from the way in which the third
parameter of the model is determined. In Ref.\citep{Osipov:2006js} the rather
sensitive value of $a_{2}^{2}$ was used, while here we choose to fix the
chiral condensate.

In our scenarios which include wavefunction renormalization we have fixed
$Z(0)=0.7$. As it can be seen in Fig.\ref{Fig1}, however, for small values of
$p$ the errors in the lattice results are rather large. Thus, $Z(0)$ is not
well constrained by lattice calculations. In order to test the sensitivity of
our results to this kind of uncertainties we have reduced it to $Z(0)=0.6$,
and considered the scenario S2 for two alternative situations. In the first
case we varied the model parameters so that $f_{\pi}$ and $m_{\pi}$ remain at
their empirical values, while in the second case we kept the model parameters
fixed. In both cases we found that most of our results change by less that
10\%, the most notorious exception being the $\pi\pi$ scattering length
$a_{2}^{2}$ which changes about 15 \%. It is interesting to note that in the
second case the pion mass and decay constant, as well as the chiral
condensate, get reduced. We obtain $m_{\pi}=138.7\ MeV$, $f_{\pi}=91.2\ MeV$
and $-<\bar{q}q>^{1/3}=323\ MeV$.

\section{Comparison with Chiral Perturbation Theory}

In the previous section we have focused our attention on the ability of our
quark model to reproduce the phenomenological $\pi$-$\pi$ scattering
parameters. An alternative point of view (see, for example Refs.
\citep{Schuren:1991sc,Holdom:1989jb,Frank:1995yb,Yang:2002re,LlanesEstrada:2003ha,Choi:2003cz})
is to consider the quark models as the generators of the pion Chiral
Perturbation Theory ($\chi$PT) Lagrangian\citep{Gasser:1983yg}. $\chi$PT
describes the low energy physics of pions in a universal way, once the order
in the momentum and chiral breaking expansion (i.e. the order in the chiral
expansion) is specified. Different scenarios for quark models will lead to
$\chi$PT Lagrangians with different values of the so-called low energy
constants (LECs). In this section we analyze the connection between our quark
scenarios and the $\chi$PT Lagrangian up to the fourth order in the chiral expansion.

To perform this connection we introduce the pionic Lagrangian
\begin{equation}
\mathcal{L}=\mathcal{L}_{2}+\mathcal{L}_{4}\ \,, \label{ChiralExpansion}%
\end{equation}
where
\begin{align}
\mathcal{L}_{2}  &  =\frac{f^{2}}{4}\left\langle \partial_{\mu}U^{\dagger
}\ \partial^{\mu}U+U^{\dagger}\chi+\chi^{\dagger}U\right\rangle
\ \ \ \ \ ,\label{L2Chiral}\\
\mathcal{L}_{4}  &  =\ell_{1}\left\langle \partial_{\mu}U^{\dagger}%
\ \partial^{\mu}U\right\rangle ^{2}+\ell_{2}\left\langle \partial_{\mu
}U^{\dagger}\ \partial_{\nu}U\right\rangle \left\langle \partial^{\mu
}U^{\dagger}\ \partial^{\nu}U\right\rangle ^{2}+\ell_{3}\left\langle \chi
U\right\rangle ^{2}+\ell_{4}\left\langle \partial_{\mu}\chi\ \partial^{\mu
}U\right\rangle ^{2}+...\ \ \,\,\ , \label{L4Chiral}%
\end{align}
and
\begin{align}
U=\exp\left(  i\frac{\vec{\tau}.\vec{\pi}}{f}\right)  \qquad;\qquad\chi
=m^{2}\left(
\begin{array}
[c]{rr}%
1 & 0\\
0 & 1
\end{array}
\right)  \ \ \ . \label{MassMatrix}%
\end{align}
Note that among all possible $O(4)$ terms only those relevant for $\pi$-$\pi$
scattering to that order have been explicitly given. To the order we are
working, the parameters $f$ and $m$ can be related with the predicted values
for $f_{\pi}$ and $m_{\pi}$ through
\begin{align}
f_{\pi}  &  = f\left(  1+\left(  \frac{m_{\pi}}{f_{\pi}}\right)  ^{2}\ell
_{4}\right)  \ ,\label{FChPTequivalente}\\
m_{\pi}^{2}  &  = m^{2}\left(  1+2\left(  \frac{m_{\pi}}{f_{\pi}}\right)
^{2}\ell_{3}\right)  \ . \label{MChPTequivalente}%
\end{align}
Using (\ref{FChPTequivalente}) and (\ref{MChPTequivalente}), we can express
the scattering parameters resulting from Eq.(\ref{ChiralExpansion}) in terms
of the $\ell_{i}$ coupling constants and the $m_{\pi}$ and $f_{\pi}$ values as
follows
\begin{align}
m_{\pi}\ a_{0}^{0}  &  = \frac{7}{32\pi}\left(  \frac{m_{\pi}}{f_{\pi}%
}\right)  ^{2}\left\{  1+\frac{1}{7}\left(  \frac{m_{\pi}}{f_{\pi}}\right)
^{2}\left[  40\ \ell_{1}+40\ \ell_{2}+10\ \ell_{3}+14\ \ell_{4}\right]
\right\} \nonumber\\
m_{\pi}\ a_{0}^{2}  &  = -\ \frac{1}{16\pi}\left(  \frac{m_{\pi}}{f_{\pi}%
}\right)  ^{2}\left\{  1-\left(  \frac{m_{\pi}}{f_{\pi}}\right)  ^{2}\left[
8\ \ell_{1}+8\ \ell_{2}+2\ \ell_{3}-2\ \ell_{4}\right]  \right\} \nonumber\\
m_{\pi}^{3}\ b_{0}^{0}  &  = \frac{1}{4\pi}\left(  \frac{m_{\pi}}{f_{\pi}%
}\right)  ^{2}\left\{  1+\frac{1}{4}\left(  \frac{m_{\pi}}{f_{\pi}}\right)
^{2}\left[  64\ \ell_{1}+48\ \ell_{2}+8\ \ell_{4}\right]  \right\} \nonumber\\
m_{\pi}^{3}\ b_{0}^{2}  &  = - \frac{1}{8\pi}\left(  \frac{m_{\pi}}{f_{\pi}%
}\right)  ^{2}\left\{  1-\frac{1}{2}\left(  \frac{m_{\pi}}{f_{\pi}}\right)
^{2}\left[  16\ \ell_{1}+24\ \ell_{2}-4\ \ell_{4}\right]  \right\} \nonumber\\
m_{\pi}^{3}\ a_{1}^{1}  &  = \frac{1}{24\pi}\left(  \frac{m_{\pi}}{f_{\pi}%
}\right)  ^{2}\left\{  1+\left(  \frac{m_{\pi}}{f_{\pi}}\right)  ^{2}\left[
-8\ \ell_{1}+4\ \ell_{2}+2\ \ell_{4}\right]  \right\} \nonumber\\
m_{\pi}^{5}\ b_{1}^{1}  &  = \frac{1}{6\pi}\left(  \frac{m_{\pi}}{f_{\pi}%
}\right)  ^{4}\left\{  -2\ \ell_{1}+\ell_{2}\right\} \nonumber\\
m_{\pi}^{5}\ a_{2}^{0}  &  = \frac{1}{15\pi}\left(  \frac{m_{\pi}}{f_{\pi}%
}\right)  ^{4}\left\{  \ell_{1}+2\ \ell_{2}\right\} \nonumber\\
m_{\pi}^{5}\ a_{2}^{2}  &  = \frac{1}{30\pi}\left(  \frac{m_{\pi}}{f_{\pi}%
}\right)  ^{4}\left\{  2\ \ell_{1}+\ell_{2}\right\}
\label{ParamScattChPTequivalente}%
\end{align}
As already mentioned, the Lagrangian (\ref{ChiralExpansion}) is valid up to
fourth order in the chiral expansion, therefore it can be fully equivalent to
our quark model scenarios only when they are treated to the same order of
approximation. Thus, to extract the LECs defining the pionic Lagrangian from
the values of the scattering parameters, $f_{\pi}$ and $m_{\pi}$ obtained in
each of our quark scenarios we should analyze the values of these parameters
for small values of $m_{c}$. In fact, we have verified that close to the
chiral limit the scattering parameters display, as a function of $\left(
m_{\pi}/f_{\pi}\right)  ^{2}$, the quadratic behavior expected from
Eqs.(\ref{ParamScattChPTequivalente}). From the determination of the
corresponding linear and quadratic coefficients it is possible to obtain the
numerical values of LECs $\ell_{i}$. It should be noticed that this procedure
for obtaining $\ell_{i}$ is completely equivalent to the bosonization of the
quark Lagrangian followed by a covariant gradient expansion (see
Ref.\citep{Schuren:1991sc} for the application of such method to the NJL
model). At this stage we are connecting our quark model at the one loop level
to the pionic Lagrangian at the tree level. Our next step is to make the
connection with the $\chi$PT Lagrangian. The main difference between the
Lagrangian (\ref{ChiralExpansion}) and the $\chi$PT Lagrangian is that our
$\ell_{i}$ parameters are finite and no pion loop contribution is present. The
scattering parameters in $\chi$PT \citep{Gasser:1983yg} include pion loop
contributions, and are written in terms of renormalized LECs $\ell_{i}^{r}$.
As expected, Eqs.(\ref{ParamScattChPTequivalente}) coincide with the ones
obtained from $\chi$PT if the corresponding pion loop contributions are
neglected. In this approximation the coupling constants $\ell_{i}$ can be
identified with the $\ell_{i}^{r}$ constants at some given renormalization
scale $\mu$.

Eqs.(\ref{ParamScattChPTequivalente}) imply several relations between the
scattering parameters. We focus on two of them
\begin{equation}
Test1=m_{\pi}\ \left(  2\ a_{0}^{0}-5\ a_{0}^{2}\right)  +m_{\pi}^{3}\ \left(
-\frac{9}{2}a_{1}^{1}-b_{0}^{0}+\frac{5}{2}b_{0}^{2}\right)  =\left\{
\begin{array}
[c]{cc}%
0 & \text{ \ \ using (\ref{ParamScattChPTequivalente})}\\
\frac{m_{\pi}^{4}}{16\pi^{4}f_{\pi}^{4}}\frac{17\pi}{12} & \text{\ \ using
}\chi\text{PT}%
\end{array}
\right.  \label{Test1}%
\end{equation}
\begin{equation}
Test2=\frac{5}{2}\ m_{\pi}\ a_{2}^{2}+\frac{3}{10}\ m_{\pi}\ b_{1}^{1}-m_{\pi
}\ a_{2}^{0}=\left\{
\begin{array}
[c]{cc}%
0 & \text{ \ \ using (\ref{ParamScattChPTequivalente})}\\
\frac{1}{16\pi^{4}f_{\pi}^{4}}\frac{7\pi}{450} & \text{\ \ using }%
\chi\text{PT}%
\end{array}
\right.  \label{Test2}%
\end{equation}
Obviously these two relations vanish when we use our pionic Lagrangian
(\ref{ChiralExpansion}) at the tree level. Therefore, a non-vanishing value
obtained for these two quantities in any other calculation must be originated
by loop corrections or by higher order terms in the chiral expansion. As
indicated in Eqs.(\ref{Test1},\ref{Test2}), in the case of the $\chi$PT
Lagrangian both relations have corrections from pionic loops. On the other
hand, the deviation from zero of $Test1$ and $Test2$ when evaluated using the
scattering parameters obtained in our quark scenarios at the physical value of
$m_{\pi}$ is originated by higher order terms in the chiral expansion. In
Table \ref{TableTest} we show the results for these two relations in our
scenarios. Also indicated are the $\chi$PT Lagrangian results, which
corresponds to the pion loop contribution of the order $\left(  m_{\pi}/2\pi
f_{\pi}\right)  ^{4}$. From this table we observe that the quark scenarios
previously studied give results for $Test1$ and $Test2$ which are of the same
order of magnitude than the pion loop contributions. This implies that the
studied quark models include higher order contributions (i.e. O(6) or higher)
which are as important as the chiral loops. The effect of these higher order
contributions is more important for the scenario S2, due to its different
behavior for large momenta.

In Table V we give the $\ell_{i}$ values corresponding to our different
scenarios. It is interesting to note that in the case of $\ell_{1}$ the listed
values result, in all cases, from an important cancellation between the box
and the sigma contributions. For $\ell_{2}$ only box contribution is present
since no scalar meson contribution is possible \citep{Ecker:1988te}. Also
given in Table V are the values of the renormalized LECs $\ell_{i}^{r}\left(
\mu\right)  $ obtained\citep{Colangelo:2001df} in the framework of $\chi$PT at
some particular values of renormalization scale $\mu$\cite{footnote}. We
observe that the sign and order of magnitude of the most accurately known LECs
$\ell_{2}^{r}$ and $\ell_{4}^{r}$ are well reproduced for small values of
$\mu$. In fact, in the case of the scenarios S1 and S2 the agreement is
remarkable good for $\mu$ around $2\ m_{\pi}$ which is a reasonable scale
since we have integrated out degrees of freedom from below the sigma mass. In
the case of the LECs $\ell_{1}^{r}$ and $\ell_{3}^{r}$, even though it is not
so good in the case of S2, the agreement is still acceptable given the
existing uncertainties in the determination of the empirical values. Finally,
as a reference, some typical values obtained within the local NJL model taken
from Ref.\cite{Schuren:1991sc} are also listed in Table V.

\section{Conclusions}

In this work we have analyzed the sigma meson mass and width together with the
$\pi$-$\pi$ scattering parameters in the context of non-local chiral quark
models with wave-function renormalization (WFR) term. We have considered two
types of momentum dependence for the quark interactions. The first one
(scenario S1) is based on the frequently used exponential form factors. The
second one (scenario S2) corresponds to a fit to the mass and renormalization
functions obtained in lattice calculations\citep{Parappilly:2005ei}, and gives
rise to a softer momentum dependence (e.g. at large momentum, the quark mass
decreases as $(p^{2})^{-3/2}$ instead of exponentially). In order to test the
influence of the WFR, we also considered a third scenario S3 which corresponds
to an exponential interaction but where this renormalization is absent.

Our results for the sigma mass are relatively stable, ranging from 552 MeV for
S2 to 683 MeV for S3. We observe that the coupling between the scalar term of
the standard chiral interaction and the new scalar term associated with the
WFR term reduces the value of the lower sigma mass, as it must be expected.
Comparing the S3 and S1 results we observe a reduction of a 10\%, while in the
case of the S2 interaction there is a further reduction of 10\% originated by
the softer momentum dependence of the interaction. The width of the sigma
follows the same reduction as its mass, as one goes from one scenario to
another. These results are less dependent on the parameterization than in the
standard NJL model. The predicted mass and width are reasonable close to the
recently reported empirical values \citep{Aitala:2000xu,Wu:2001vz}.

Regarding the $\pi-\pi$ scattering parameters, we have compared our results
with the phenomenological determination made in Ref.\citep{Kaminski:2006qe}.
Although the existence of the chiral limit relations,
Eqs.(\ref{ScatLengthChiral}), for the S- and P-wave scattering length and
slope parameters reduces the sensitivity of these parameters to the choice of
the different quark interactions, we have been able to discriminate between
these interactions by going to higher order in the momentum expansion or to
higher partial waves. We conclude that although the exponential interaction is
able to reproduce the scattering lengths parameters rather well the
description of higher power coefficients turns out to be, in general, less
accurate as the power in $q^{2}$ increases. This can be clearly seen in the
case of higher partial waves. On the other hand, the momentum dependence of
the scenario S2 seems to be better adapted for the description of the existing
empirical data. Comparing the predictions of the scenarios based on
exponential interactions, S1 and S3, we observe that the presence of the WFR
term tends to improve the results, even though its effect is less noticeable
that the one produced by the difference on the momentum dependence of the interactions.

Finally, we have analyzed the relation of our quark scenarios with the chiral
Lagrangian up to $O(4)$ in the chiral expansion. In particular, we have
obtained predictions for the low energy constants $\ell_{i}$ involved in $\pi
$-$\pi$ scattering within our scenarios. The procedure we followed, using the
scattering parameters, is equivalent to the standard method of bosonization
followed by covariant gradient expansion. Our predicted values for $\ell_{i}$
are in relative good agreement with the values for the renormalized $\ell
_{i}^{r}$ constants defined in the $\chi$PT calculations \citep{Gasser:1983yg}
for a $\mu$ value about $2m_{\pi}$. They are also in the range of values
obtained in the NJL model calculation of Ref. \citep{Schuren:1991sc}. We have
been able to define combinations of the scattering parameters which allow to
discriminate between higher chiral corrections (O(6) or higher) and pion loop
corrections. We observe that the higher order corrections included in our
non-local quark model calculations at physical $m_{\pi}$ are of the same order
that the pion loop corrections not considered in this work. The effect of such
corrections in our predictions for the mesonic observables is an issue that
deserves further investigation.

\section*{Acknowledgements}

The authors wish to thank D. Gomez Dumm and J. Portoles for useful
discussions. NNS acknowledges the support of CONICET (Argentina) grant PIP
6084, and ANPCyT (Argentina) grant PICT04 03-25374, and SN the support of the
Sixth Framework Program of the European Commision under the Contract No.
506078 (I3 Hadron Physics) and MEC (Spain) under the Contract FPA
2007-65748-C02-01 and EU FEDER.




\vspace*{3cm}

\begin{table}[h]
\caption{Model parameters and results for some alternative parameterizations.}%
\label{tab1}%
\vspace*{0.5cm}
\par
\begin{centering}
\begin{tabular}{ccccc}
\hline
&  & \hspace{.5cm} S1 \hspace{.5cm}  & \hspace{.5cm} S2 \hspace{.5cm}  & S3\tabularnewline
\hline
$m_{c}$  & MeV  & 5.70  & 2.37  & 5.78\tabularnewline
$G_{s}\Lambda_{0}^{2}$  &  & 32.03  & 20.82  & 20.65\tabularnewline
$\Lambda_{0}$  & MeV  & 814.42  & 850.00  & 752.2\tabularnewline
$\varkappa_{P}$  & GeV  & 4.18  & 6.03  & $-$\tabularnewline
$\Lambda_{1}$  & MeV  & 1034.5  & 1400  & $-$\tabularnewline
\hline
$\bar{\sigma}_{1}$  & MeV  & 529  & 442  & 424\tabularnewline
$\bar{\sigma}_{2}$  &  & -0.43  & -0.43  & $-$\tabularnewline
$M(0)$  & MeV  & 375  & 311  & 430\tabularnewline
$Z(0)$  &  & 0.7  & 0.7  & 1.0\tabularnewline
$-<q\bar{q}>^{1/3}$  & MeV  & 240  & 326  & 240\tabularnewline
\hline
$m_{\pi}$  & MeV  & 139  & 139  & 139\tabularnewline
$g_{\pi q\bar{q}}$  &  & 5.74  & 4.74  & 4.62\tabularnewline
$f_{\pi}$  & MeV  & 92.4  & 92.4  & 92.4\tabularnewline
\hline
$m_{\sigma}$  & MeV  & 622  & 552  & 683\tabularnewline
$g_{\sigma q\bar{q}}^{(0)}$  &  & 5.97  & 4.60  & 5.08\tabularnewline
$g_{\sigma q\bar{q}}^{(1)}$  &  & $-0.77$  & $-0.26$  & $-$\tabularnewline
$\Gamma_{\sigma\pi\pi}$  & MeV  & 263  & 182  & 347\tabularnewline
\hline
\end{tabular}
\par\end{centering}
\end{table}

\begin{table}[h]
\caption{$\pi-\pi$ scattering parameters for S and P waves.}%
\label{tab2}%
\vspace*{0.5cm}
\par
\begin{centering}
\begin{tabular}{cccccccc}
\hline
& contribution  & \hspace{0.5cm} S1 \hspace{0.5cm}  & \hspace{0.5cm} S2 \hspace{0.5cm}  & \hspace{0.5cm} S3 \hspace{0.5cm}  & \multicolumn{2}{c}{NJL} & Empirical\tabularnewline
&  &  &  &  & Ref.\citep{Schuren:1991sc}  & Ref.\citep{Bernard:1992mp}  & Ref.\citep{Kaminski:2006qe}\tabularnewline
\hline
$(m_{\pi})\times a_{0}^{0}$  & box  & $-1.536$  & $-1.279$  & $-1.618$  &  &  & \tabularnewline
& $\sigma$  & $1.718$  & $1.470$  & $1.798$  &  &  & \tabularnewline
\hline
& Total  & $0.182$  & $0.191$  & $0.180$  & $0.18$  & $0.19$  & $0.223\pm0.009$\tabularnewline
\hline
$(m_{\pi}^{3})\times b_{0}^{0}$  & box  & $0.114$  & $0.117$  & $0.114$  &  &  & \tabularnewline
& $\sigma$  & $0.116$  & $0.146$  & $0.107$  &  &  & \tabularnewline
\hline
& Total  & $0.230$  & $0.263$  & $0.221$  & $0.22$  & $0.27$  & $0.290\pm0.006$\tabularnewline
\hline
$(m_{\pi}^{5})\times c_{0}^{0}$  & box  & $-0.0086$  & $0.0233$  & $-0.0076$  &  &  & \tabularnewline
& $\sigma$  & $0.0412$  & $0.0663$  & $0.0302$  &  &  & \tabularnewline
\hline
& Total  & $0.0326$  & $0.0897$  & $0.0226$  &  &  & \tabularnewline
\hline
$(m_{\pi}^{7})\times d_{0}^{0}$  & box  & $0.0005$  & $0.065$  & $0.0004$  &  &  & \tabularnewline
& $\sigma$  & $0.0087$  & $0.019$  & $0.0051$  &  &  & \tabularnewline
\hline
& Total  & $0.0092$  & $0.085$  & $0.0055$  &  &  & \tabularnewline
\hline
\hline
$(m_{\pi})\times a_{0}^{2}$  & box  & $-0.6851$  & $-0.5790$  & $-0.7170$  &  &  & \tabularnewline
& $\sigma$  & $0.6404$  & $0.5346$  & $0.6721$  &  &  & \tabularnewline
\hline
& Total  & $-0.0447$  & $-0.0444$  & $-0.0449$  & $-0.046$  & $-0.044$  & $-0.0444\pm0.0045$\tabularnewline
\hline
$(m_{\pi}^{3})\times b_{0}^{2}$  & box  & $-0.053$  & $-0.049$  & $-0.051$  &  &  & \tabularnewline
& $\sigma$  & $-0.031$  & $-0.034$  & $-0.033$  &  &  & \tabularnewline
\hline
& Total  & $-0.084$  & $-0.083$  & $-0.084$  & $-0.091$  & $-0.079$  & $-0.081\pm0.003$\tabularnewline
\hline
$(m_{\pi}^{5})\times c_{0}^{2}$  & box  & $0.0080$  & $0.0078$  & $0.0082$  &  &  & \tabularnewline
& $\sigma$  & $0.0042$  & $0.0056$  & $0.0034$  &  &  & \tabularnewline
\hline
& Total  & $0.0121$  & $0.0134$  & $0.0116$  &  &  & \tabularnewline
\hline
$(m_{\pi}^{7})\times d_{0}^{2}$  & box  & $-0.0005$  & $-0.0006$  & $-0.0005$  &  &  & \tabularnewline
& $\sigma$  & $-0.0006$  & $-0.0011$  & $-0.0004$  &  &  & \tabularnewline
\hline
& Total  & $-0.0011$  & $-0.0017$  & $-0.0009$  &  &  & \tabularnewline
\hline
\hline
$m_{\pi}\times(2a_{0}^{0}+7a_{0}^{2})$  &  & 0.052  & 0.072  & 0.046  & $0.04$  & 0.072  & $0.135\pm0.036$\tabularnewline
\hline
\hline
$(m_{\pi}^{3}\ 10^{3})\times a_{1}^{1}$  & box  & $25.1$  & $23.9$  & $24.7$  &  &  & \tabularnewline
& $\sigma$  & $10.5$  & $11.3$  & $11.1$  &  &  & \tabularnewline
\hline
& Total  & $35.7$  & $35.2$  & $35.7$  & $37$  & $34$  & $38.1\pm0.9$\tabularnewline
\hline
$(m_{\pi}^{5}\ 10^{3})\times b_{1}^{1}$  & box  & $5.56$  & $4.60$  & $5.34$  &  &  & \tabularnewline
& $\sigma$  & $-2.10$  & $-2.85$  & $-1.72$  &  &  & \tabularnewline
\hline
& Total  & $3.45$  & $1.75$  & $3.62$  &  &  & $5.13\pm0.15$\tabularnewline
\hline
$(m_{\pi}^{7}\ 10^{3})\times c_{1}^{1}$  & box  & $0.21$  & $-2.70$  & $0.15$  &  &  & \tabularnewline
& $\sigma$  & $0.38$  & $0.63$  & $0.25$  &  &  & \tabularnewline
\hline
& Total  & $0.59$  & $-2.06$  & $0.40$  &  &  & \tabularnewline
\hline
\end{tabular}
\par\end{centering}
\end{table}

\begin{table}[h]
\caption{Scattering lengths and slope parameters for D and F waves.}%
\label{tab3}%
\vspace*{0.5cm}
\par
\begin{centering}
\begin{tabular}{cccccccc}
\hline
& contribution  & \hspace{0.5cm} S1 \hspace{0.5cm}  & \hspace{0.5cm} S2 \hspace{0.5cm}  & \hspace{0.5cm} S3 \hspace{0.5cm}  & \multicolumn{2}{c}{NJL} & Empirical\tabularnewline
&  &  &  &  & Ref.\citep{Schuren:1991sc}  & Ref.\citep{Bernard:1992mp}  & Ref.\citep{Kaminski:2006qe}\tabularnewline
\hline
\hline
$(m_{\pi}^{5}\ 10^{4})\times a_{2}^{0}$  & box  & $9.71$  & $9.76$  & $9.93$  &  &  & \tabularnewline
& $\sigma$  & $4.20$  & $5.67$  & $3.44$  &  &  & \tabularnewline
\hline
& Total  & $13.91$  & $15.43$  & $13.37$  & $13.7$  & $16.7$  & $18.33\pm0.36$\tabularnewline
\hline
$(m_{\pi}^{7}\ 10^{4})\times b_{2}^{0}$  & box  & $0.98$  & $0.85$  & $1.04$  &  &  & \tabularnewline
& $\sigma$  & $-1.28$  & $-2.20$  & $-0.86$  &  &  & \tabularnewline
\hline
& Total  & $-0.30$  & $-1.34$  & $0.18$  &  &  & $-3.82\pm0.25$\tabularnewline
\hline
\hline
$(m_{\pi}^{5}\ 10^{4})\times a_{2}^{2}$  & box  & $-2.74$  & $-2.95$  & $-2.43$  &  &  & \tabularnewline
& $\sigma$  & $4.20$  & $5.67$  & $3.44$  &  &  & \tabularnewline
\hline
& Total  & $1.46$  & $2.72$  & $1.01$  & $1.1$  & $3.2$  & $2.46\pm0.25$\tabularnewline
\hline
$(m_{\pi}^{7}\ 10^{4})\times b_{2}^{2}$  & box  & $0.08$  & $0.07$  & $0.13$  &  &  & \tabularnewline
& $\sigma$  & $-1.28$  & $-2.20$  & $-0.86$  &  &  & \tabularnewline
\hline
& Total  & $-1.19$  & $-2.14$  & $-0.73$  &  &  & $-3.59\pm0.18$\tabularnewline
\hline
\hline
$(m_{\pi}^{7}\ 10^{5})\times a_{3}^{1}$  & box  & $0.82$  & $1.15$  & $0.7$  &  &  & \tabularnewline
& $\sigma$  & $1.82$  & $3.09$  & $1.2$  &  &  & \tabularnewline
\hline
& Total  & $2.65$  & $4.24$  & $1.9$  &  &  & $6.05\pm0.29$\tabularnewline
\hline
$(m_{\pi}^{9}\ 10^{5})\times b_{3}^{1}$  & box  & $0.06$  & $0.0$  & $0.07$  &  &  & \tabularnewline
& $\sigma$  & $-0.70$  & $-1.6$  & $-0.40$  &  &  & \tabularnewline
\hline
& Total  & $-0.64$  & $-1.6$  & $-0.33$  &  &  & $-4.41\pm0.36$\tabularnewline
\hline
\end{tabular}
\par\end{centering}
\end{table}

\pagebreak

\begin{table}[t]
\caption{Results for $Test1$ and $Test2$ defined in Eqs.(\ref{Test1}%
,\ref{Test2}) as obtained in our quark scenarios (S1,S2 and S3) and Chiral
Perturbation Theory to O(4) ($\chi$PT). The results obtained using the
empirical values of Ref.\citep{Kaminski:2006qe} (Empirical) are also given. }%
\label{TableTest}%
\vspace*{0.5cm}
\par
\begin{centering}
$\hspace*{-2cm}\begin{tabular}{cccccc}
\hline   &  S1  &  S2  &  S3  &  $~~~~\chi$PT$~~~~$  &  Empirical \cite{Kaminski:2006qe}\\
\hline $Test1\times10^{2}$  &  $-1.2$  &  $-2.5$  &  $-1.1$  &  $1.5$  &  $0.40\pm4.4$\\
$Test2\times\left(m_{\pi}^{4}\ 10^{4}\right)$  &  $~~0.1~~$  &  $~~-3.3~~$  &  $~~0.006~~$  &  $1.6$  &  $3.21\pm1.4$\\\hline \end{tabular}\ \ $
\par\end{centering}
\end{table}

\begin{table}[ptb]
\caption{Values of $\ell_{i}$ obtained in our different scenarios. The $\chi
$PT values of $\ell_{i}^{r}$ as a funtion of $\mu$ are obtained from
Ref.\citep{Colangelo:2001df}. The last two columns corresponds to the NJL
predictions from Ref. \cite{Schuren:1991sc} for two different constituent
quark mass: $M=220,$264~MeV.}%
\label{Tablel}%
\vspace*{0.5cm}
\par
\par
\begin{centering}
\begin{tabular}{lccccccccccc}
\hline
& & \multicolumn{3}{c}{Non Local Quark Model } & & \multicolumn{3}{c}{$\chi$PT ($\ell_{i}^{r}(\mu)$)} & & \multicolumn{2}{c}{NJL}
\tabularnewline \cline{3-5} \cline{7-9} \cline{11-12}
& & \ \ S1  & \ \ S2  & \ \hspace*{0.1cm} \ \ S3 \hspace*{0.5cm}  & & $\ \ \ \mu=m_{\rho}$  & $\ \ \ \mu=2\ m_{\pi}$  & $\ \ \mu=m_{\pi}$  & & \ \ \ $M=220$  & $M=264$ \tabularnewline
\hline
$\ell_{1}\times10^{3}$ \hspace*{0.5cm}  & & \multicolumn{1}{r}{$\ -1.39$} & \multicolumn{1}{r}{$\ \ \ 0.26\ $} &
\multicolumn{1}{c}{$\ -2.07\hspace*{0.5cm}$} & &
\multicolumn{1}{r}{$\ \ -4.0\pm0.6\ \ $} & \multicolumn{1}{r}{$\ \ -1.9\pm0.6\ \ $} & \multicolumn{1}{r}{$\ \ -0.4\pm0.6\ $} & &
$-0.63$  & $-2.28$\tabularnewline
$\ell_{2}\times10^{3}$  & & \multicolumn{1}{r}{$6.46$} & \multicolumn{1}{r}{$6.41\ $} & \multicolumn{1}{c}{$6.51\hspace*{0.5cm}$} & &
\multicolumn{1}{r}{$1.9\pm0.2\ \ $} & \multicolumn{1}{r}{$6.2\pm0.2\ \ $} & \multicolumn{1}{c}{$9.1\pm0.2\ $} & &
$6.29$  & $6.18$\tabularnewline
$\ell_{3}\times10^{3}$ & & \multicolumn{1}{r}{$-2.3$} & \multicolumn{1}{r}{$-4.1\ $} & \multicolumn{1}{c}{$-1.1\hspace*{0.5cm}$} & &
\multicolumn{1}{r}{$1.5\pm4.0\ \ $} & \multicolumn{1}{r}{$-1.8\pm4.0\ \ $} & \multicolumn{1}{c}{$-4.0\pm4.0\ $} & &
$-8.50$  & $-3.48$\tabularnewline
$\ell_{4}\times10^{3}$ & & \multicolumn{1}{r}{$17.2$} & \multicolumn{1}{r}{$20.3\ $} & \multicolumn{1}{c}{$15.0\hspace*{0.5cm}$} & &
\multicolumn{1}{r}{$6.2\pm1.3\ \ $} & \multicolumn{1}{r}{$19.1\pm1.3\ \ $} & \multicolumn{1}{c}{$27.9\pm1.3\ $} & &
$22.73$  & $12.16$\tabularnewline
\hline
\end{tabular}
\par\end{centering}
\end{table}

\begin{thebibliography}{99}
\bibitem {footnote}It should be noticed that $f_{\pi}$ and $m_{\pi}$ are also
affected by the pion loops. For $\mu=2m_{\pi}$ the pion loops produce a factor
1.02 for $f_{\pi}$ and a factor 0.99 for $m_{\pi}^{2}$. This effect has been
neglected in the present discussion.
\end{thebibliography}
\end{document}